\documentclass[aip,apl,twocolumn,showpacs,amsmath,amssymb,a4paper,floatfix,reprint, groupedaddress]{revtex4-1}
\usepackage{mathptmx}
\usepackage[utf8]{inputenc} 
\usepackage{graphicx}
\usepackage{xspace}
\usepackage{epstopdf}
\usepackage{subfigure}
\usepackage[
,textwidth=17.4cm
,textheight=25.3cm
,verbose
,dvips
]{geometry}

\newcommand{\celsius}{$^{\circ}$C\xspace}

\begin{document}


\title{All-electrical spin injection and detection in the Co$_2$FeSi/GaAs hybrid system in the local and non-local configuration} 



\author{P. Bruski}
\email{bruski@pdi-berlin.de}
\author{Y. Manzke}
\author{R. Farshchi}
\author{O. Brandt}
\author{J. Herfort}
\author{M. Ramsteiner}
\affiliation{Paul-Drude-Institut f\"ur Festk\" orperelektronik,
Hausvogteiplatz 5--7, 10117 Berlin, Germany}
\date{\today}

\begin{abstract}
We demonstrate the electrical injection and detection of spin-polarized electrons in the Co$_2$FeSi/GaAs hybrid system using lateral transport structures. Spin valve signatures and characteristic Hanle curves are observed both in the non-local and the local configuration. 
The comparatively large magnitude of the local spin valve signal and the high signal-to-noise ratio are attributed to the large spin polarization at the Fermi energy of Co$_2$FeSi in the well-ordered \textit{L}2$_1$ phase.
\end{abstract}


\maketitle 


Most spin-based semiconductor devices proposed to date rely on the ability to inject, transport, manipulate and detect spin-polarized carriers by purely electrical means.\cite{DattaDas1989, Sugahara2004}
As a fundamental proof of the all-electrical injection and detection of spins in a lateral device structure, non-local (NL) spin valve measurements with separated charge and spin currents are most appropriate.\cite{Johnson1985, Jedema2001} A spin accumulation generated in the transport channel is probed by a detector contact placed outside the current path. The detector measures an electrical signal that is purely spin related.

However, NL spin detection is not sufficient for operational spintronic devices that require an electrical signal in the local (L) configuration, i.e., an electrical spin signal resulting from a spin-polarized charge current flowing between a source and a drain contact. As a matter of fact, this kind of local spin valve operation is experimentally much more difficult to achieve than the NL one.
The difficulty of the corresponding 2-point-arrangement is caused by the large electrical background signal and a strong contribution of the (not spin-related) contact resistances as well as parasitic effects.\cite{Tang2002} In order to minimize their influence, high spin injection and detection efficiencies are essential. Therefore a proper choice of the injector material can be crucial. From this point of view, half-metals are the ultimate solution regarding spin injection and detection, given that they are 100\% spin polarized at the Fermi energy.
The ferromagnetic Heusler alloy Co$_2$FeSi is predicted to be a half-metal in its ordered \textit{L}2$_1$ phase\cite{Wurmehl2005, Bruski2011} and is, in addition, closely lattice matched to GaAs.\cite{Hashimoto2005} A promising spin injection efficiency of more than 50\% has been demonstrated for Co$_2$FeSi/(Al,Ga)As hybrid structures.\cite{Ramsteiner2008}
In this Letter, we study the all-electrical injection and detection of spins in the non-local and local configuration in the Co$_2$FeSi/GaAs hybrid system using a lateral device structure.

\begin{figure}[htbp!] 
\includegraphics[width=0.23\textwidth,height=0.13\textheight]{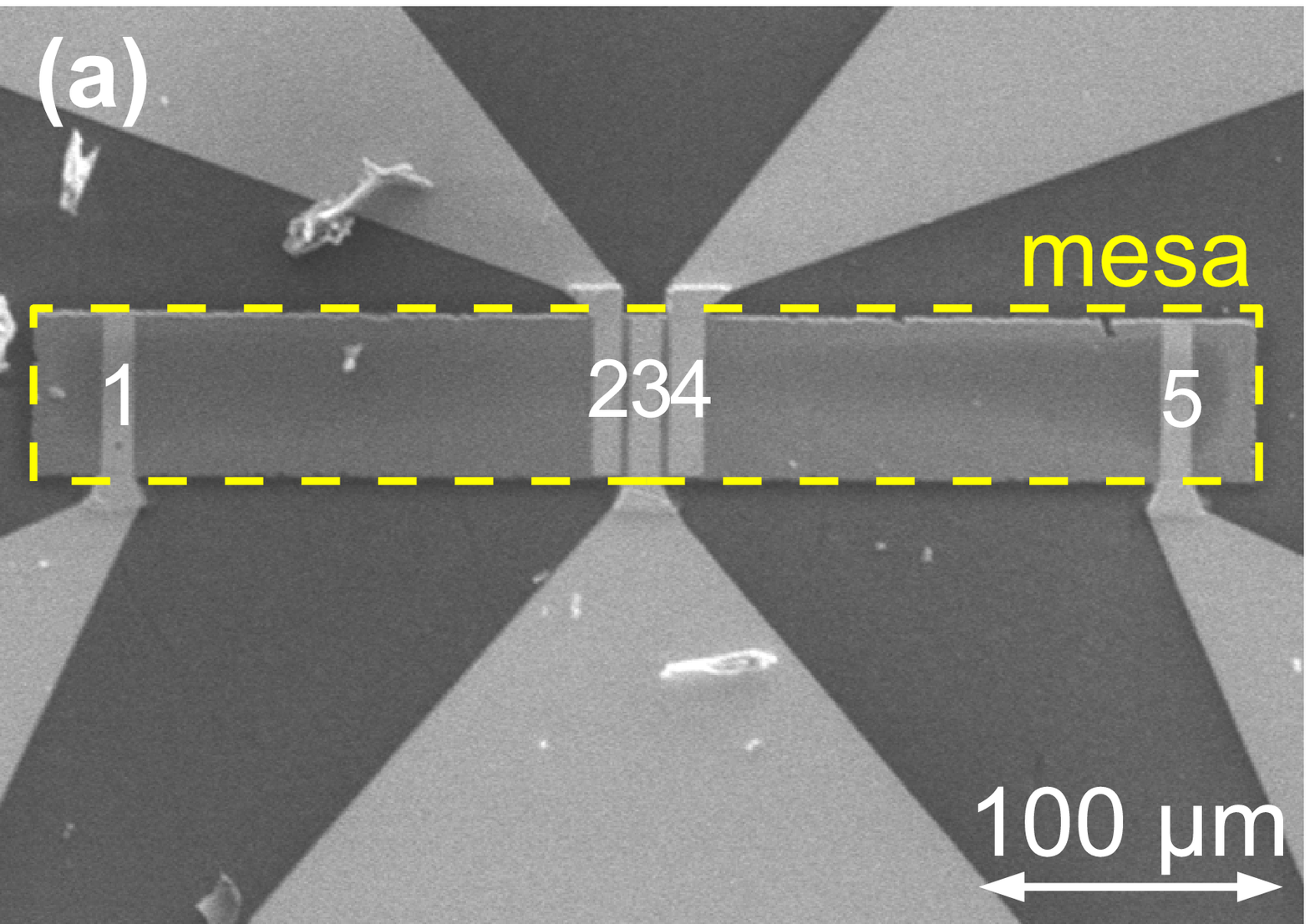}%
 \hfill
\includegraphics[width=0.25\textwidth]{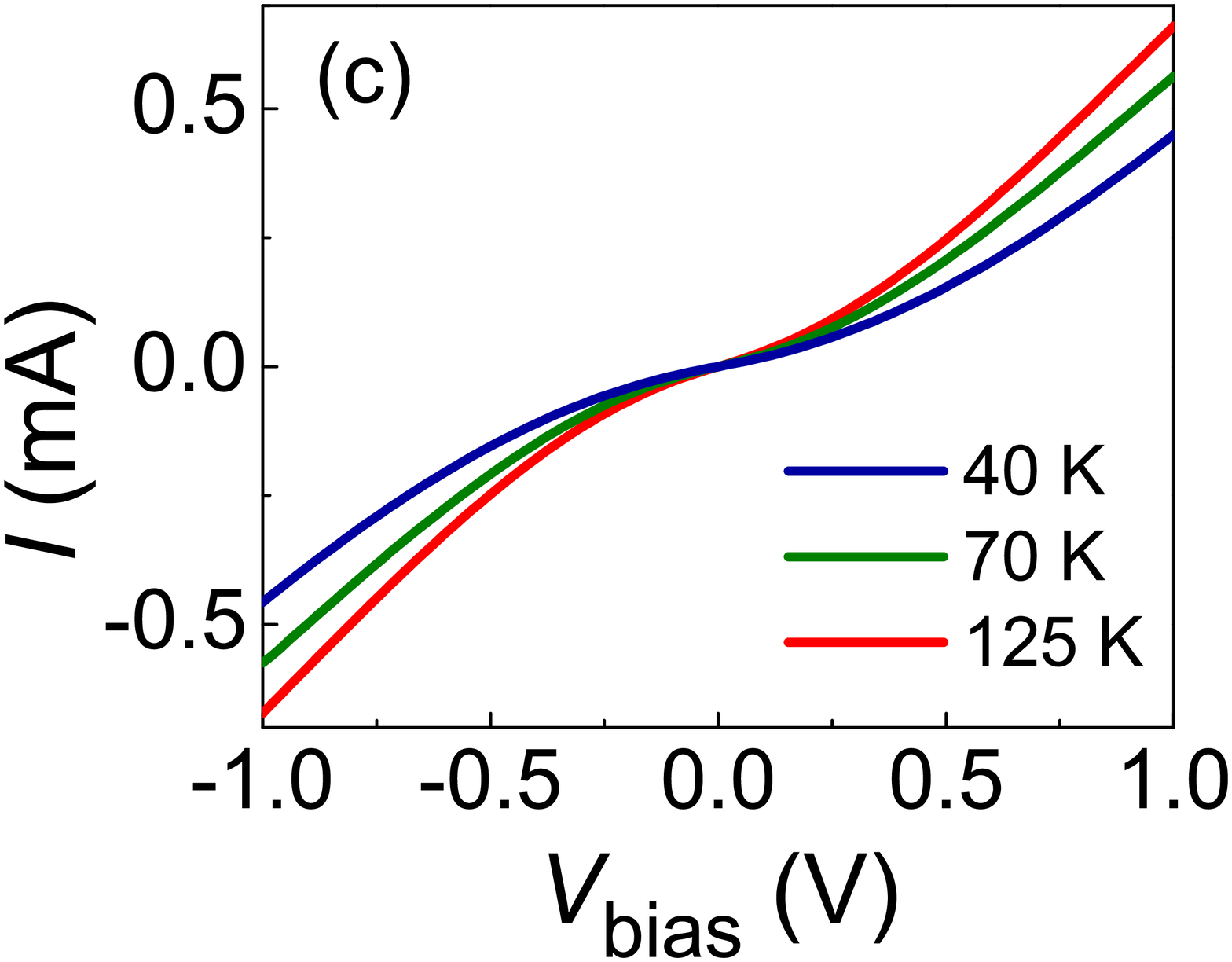}
 \\
\includegraphics[width=8cm]{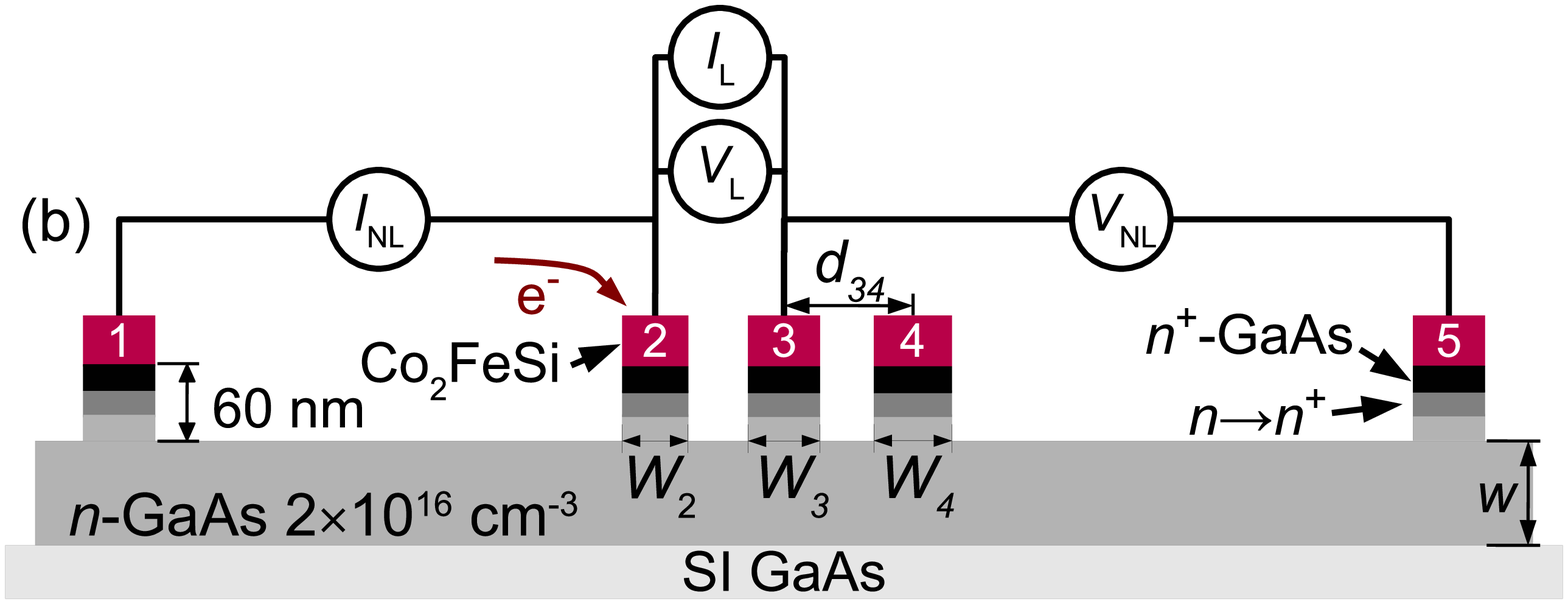}
  \caption{(a) Scanning electron micrograph of the lateral spin device for the NL voltage measurement in the top view. (b) I--V characteristics of the lateral spin device measured between contacts 1 and 2 at different temperatures. (c) Cross-sectional diagram of the lateral device geometry.}
  \label{fig1}
\end{figure}

The investigated samples were grown by molecular beam epitaxy on semi-insulating (SI) GaAs(001) substrates, processed by wet chemical etching and photolithography, and finalized by the evaporation of Au bondpads. The samples consist of a 1500~nm thick, lightly $n$-doped GaAs spin-transport layer ($2\times10^{16}$~cm$^{-3}$) followed by a 15~nm thick transition layer $n\rightarrow n^+$ and a 15~nm thick $n^+$-layer ($5\times10^{18}$~cm$^{-3}$). A 16~nm thick Co$_2$FeSi layer was deposited onto this semiconductor structure at a substrate temperature of 280\celsius. The highly $n$-doped GaAs layer directly beneath the Co$_2$FeSi forms a narrow Schottky barrier. Detailed information on the growth of the Heusler alloy Co$_2$FeSi is provided elsewhere.\cite{Hashimoto2005, Hashimoto2005a, Hashimoto2007}

A scanning electron micrograph of the lateral device structure is shown in Fig.~\ref{fig1}(a). The conductive mesa region is 400$\times 50~\mu$m$^2$ large with stripe widths $W_2$, $W_3$, and $W_4$ of 9, 10, and 11$~\mu$m, respectively. The edge-to-edge spacing between stripes 2 to 4 is 3$~\mu$m leading to center-to-center separations of $d_{23}=12.5$, $d_{34}=13.5$, and $d_{24}=26~\mu$m. The distances $d_{12}=d_{45}=145~\mu$m are much larger than the spin diffusion length.
 The measurements in the L and NL configuration were carried out on the same samples by a conventional dc method with a current of 400~$\mu$A as indicated in Fig.~\ref{fig1}(b).

To evaluate the electrical properties of the Co$_2$FeSi/GaAs Schottky contacts, we measured the two-terminal current~(I)--voltage~(V) characteristics for different contact pairs at different temperatures. The representative I--V characteristics measured between contacts~1 and 2 are shown in Fig.~\ref{fig1}(c). The curves are nonlinear at all temperatures and show a very weak temperature dependence, indicating that tunneling through the interface is dominant.\cite{Kasahara2012} An insulator-like temperature behavior of the zero bias resistance (not shown here) supports this assumption.\cite{Joensson2000}

Evidence for electrical spin injection and detection has been obtained by spin valve measurements. For these experiments, an external magnetic field ($B_{||}$) is applied parallel to the long side of the Co$_2$FeSi stripes, i.e., along the easy axis of magnetization. The measured voltage depends on the relative magnetization orientation of the injector and detector stripes. During a sweep of the external field $B_{||}$, this relative magnetization orientation changes twice from the parallel to the antiparallel condition due to slightly different coercive fields of the injector and detector caused by small variations in their stripe widths.

\begin{figure}[htb!]
\includegraphics*[width=8cm]{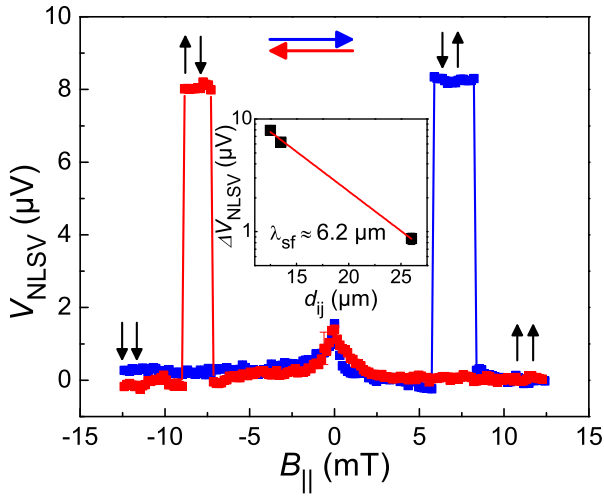}%
\caption{Non-local spin signal in the spin valve geometry as a function of an in-plane magnetic field $B_{||}$ applied along the stripes with a linear background subtracted. The peak around 0~T is induced by dynamic nuclear polarization.\cite{Salis2009b} Inset: Dependence of the difference between the NL voltage in the antiparallel and the parallel configuration $\Delta V_{\textrm{NLSV}}=V_{\uparrow \downarrow}-V_{\uparrow \uparrow}$ on the injector-detector separation $d_\textrm{ij}$ measured at a current of $I=400~\mu$A at 40~K.}
\label{fig2}
\end{figure}
In the case of the NL measurements, spin-polarized electrons are injected into the GaAs channel at stripe~2 and drift towards stripe~1. The injected spins, however, diffuse in either direction from stripe~2. While there is no charge flow between stripes~3 and 5, the diffusion-induced imbalance in the population of the two spin channels leads to a chemical potential difference. Consequently, a NL voltage can be detected between stripes~3 and 5 as a measure of the spin-injection efficiency at stripe~2 [cf.\ Fig.~\ref{fig1}(b)].

Fig.~\ref{fig2} shows the NL voltage measured in this way. The observed jumps in the voltage correspond to the switching between parallel and antiparallel magnetization of stripes~2 and 3. These characteristic spin valve signatures provide clear evidence for successful electrical injection and detection of spin polarized electrons.

At a distance $x$ from the injector, the voltage can be expressed by:\cite{Johnson1993, Jedema2003, Fabian2007}
\begin{equation}
V_{\textrm{NLSV}}=\pm \frac{P_{\textrm{inj}}P_{\textrm{det}}I\lambda_{\textrm{sf}}\rho_\textrm{N}}{2S}\exp(-x/\lambda_{\textrm{sf}}),
\label{sv}
\end{equation}
where $I$ is the bias current. $\rho_\textrm{N}$, $\lambda_{\textrm{sf}}$ and $S$ are the resistivity, spin diffusion length, and the cross-sectional area of the nonmagnetic channel, respectively. $P_\textrm{{inj(det)}}$ is the efficiency of the spin injection (detection) at the respective contact. The $+$ ($-$) sign corresponds to the parallel (antiparallel) configuration of the injector and detector electrodes.
From the dependence of the difference between the parallel and antiparallel signal $\Delta V_{\textrm{NLSV}}$ on the injector-detector separation $d_{\textrm{ij}}$ (cf.\ inset of Fig.~\ref{fig2}) we estimate the spin diffusion length in the GaAs-channel as $\lambda_{\textrm{sf}}=6.2~\mu$m. This value is in good agreement with values obtained by other groups for a similar doping of the GaAs channel.\cite{Lou2007, Ciorga2009, Salis2010}

\begin{figure}[htbp!]
\includegraphics*[width=8cm]{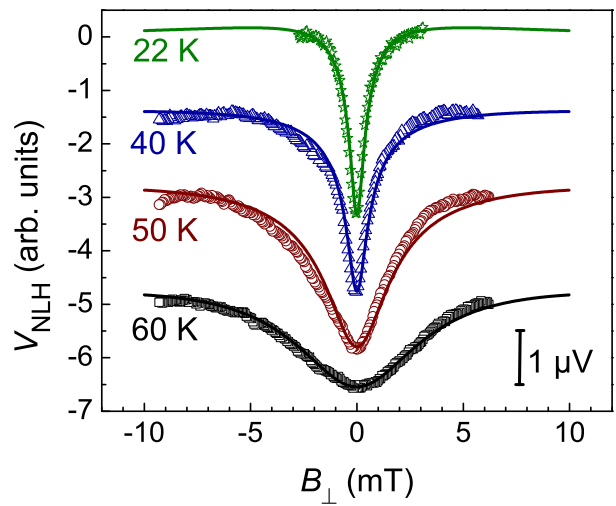}
\caption{Temperature dependence of the NL voltage $V_{\textrm{NLH}}$ (symbols) as a function of the out-of-plane magnetic field $B_\perp$. The solid lines are fits according to Eq.~(\ref{hanle}). The curves have been offset for clarity.}
\label{fig3}
\end{figure}

The most robust proof for all-electrical spin injection and detection utilizes the Hanle effect, which reveals spin precession in an external magnetic field. For the corresponding experiments, the voltage is measured in the same way as described above. The in-plane magnetization orientations of the injector and detector stripes are kept fixed in the parallel or antiparallel configuration and the external magnetic field $B_\perp$ is now applied perpendicular to the sample plane. $B_\perp$ does not influence the in-plane Co$_2$FeSi magnetization but causes a precession of the spins, which are injected into the GaAs channel. For parallel injector and detector magnetizations, the minimum voltage occurs at $B_\perp=0$ as confirmed by the measured data shown in Fig.~\ref{fig3}. For finite fields $B_\perp$, the spin precession leads to a misorientation of the spin polarization beneath the detector stripe diminishing the measured signal. The voltage in the Hanle geometry for the parallel configuration can be expressed by a one-dimensional spin drift-diffusion equation, which takes into account spin relaxation and precession, and whose solution reads:\cite{Jedema2002, Johnson1985}
\begin{equation}
V_\textrm{{NLH}}=V_0\int_{0}^\infty \, \mathrm{dt}\frac{1}{\sqrt{4\pi Dt}}e^{-d_\textrm{ij}^2/4Dt} e^{-t/\tau_\mathrm{s}}\cos(\Omega_\mathrm{L} t),
\label{hanle}
\end{equation}
with $V_0=(\pm P_{\textrm{inj}}P_{\textrm{det}}I\lambda_{\textrm{sf}}\rho_\textrm{N} /2S) (2\lambda_{\textrm{sf}}/\tau_\textrm{s})$. 
$D$ is the spin diffusion coefficient, $\tau_\textrm{s}$ the spin relaxation time and\\
 $\Omega_\textrm{L}=g\mu_B B_\perp/\hbar$ the Larmor frequency, where the electron g factor for GaAs is $g=-0.44$, $\mu_B$ is the Bohr magneton and $\hbar$ is the reduced Planck constant. 

The NL Hanle signal $V_{\textrm{NLH}}$ vs. $B_\perp$ is displayed in Fig.~\ref{fig3} for different sample temperatures. The good agreement between the measured data and the fit (cf.\ Fig.~\ref{fig3}) obtained by Eq.~(\ref{hanle}) using $D$, $\tau_\textrm{s}$, and  $V_0$ as fit parameters provides further evidence for successful all-electrical injection and detection of spin polarized electrons. 
An increase in the temperature leads to an increase in the linewidth and therefore, as expected, to a decrease in the spin lifetime from 64~ns at 22~K to 11~ns at 60~K. Both of these values are comparatively large.\cite{Lou2007,Ciorga2009} Furthermore, using $\lambda_{\textrm{sf}}=6.2~\mu$m, $I=400~\mu$A, $\rho_\textrm{N}=8.9\times 10^{-4}~\Omega$m (measured separately on the same sample using a Hall structure) and $S=75\times 10^{-12}$~m$^2$, a spin injection efficiency of $P_{\textrm{inj}}=16\%$ has been extracted by the fitting procedure. Regarding our crude assumption $P_{\textrm{inj}}=P_{\textrm{det}}$, the obtained value of $P_{\textrm{inj}}$ is in reasonable agreement with previous results obtained from Co$_2$FeSi/(Al,Ga)As spin light-emitting diodes.\cite{Ramsteiner2008}

In the case of the local measurements, the spin and charge currents are no longer separated. The measure of the spin signal in the local configuration is the magnetoresistance (MR) ratio $\Delta R/R_\textrm{P}=(R_{\textrm{AP}}-R_\textrm{P})/R_\textrm{P}$, where $R_\textrm{P}$ $(R_{\textrm{AP}})$ represents the resistance $R=V_{\textrm{L}}/I_{\textrm{L}}$ between stripes~2 and 3 [cf.\ Fig.~\ref{fig1}(b)] for the parallel (antiparallel) source and drain contact magnetizations. The requirements for a sizable spin signal in the local configuration have been theoretically discussed.\cite{Fert2001, Dery2006, Fert2007} The most crucial parameter is the ratio of $r^*_\textrm{b}$ and $r_N$. $r^*_\textrm{b}$ is the contact tunnel resistance at the interface between the ferromagnet and the semiconductor and the spin resistance $r_N$ is the product of the resistivity $\rho_\textrm{N}$ and the spin diffusion length $\lambda_{\textrm{sf}}$ within the semiconductor. A high ratio $r^*_\textrm{b}/r_\textrm{N}$ is needed to overcome the so-called conductivity mismatch\cite{Schmidt2000}, but a too high value causes the spins to relax such that it prevents their detection. As a result, a window exists for the ratio $r^*_b/r_N$ for which the obtained signal is optimal.\cite{Fert2001, Fert2007} This window is given by:
\begin{equation}
\left(\frac{W}{w}\right)\left(\frac{d_\textrm{ij}}{\lambda_{\textrm{sf}}}\right)^2\ll\frac{r^*_\textrm{b}}{r_\textrm{N}}\ll\left(\frac{W}{w}\right),
\label{window}
\end{equation}
where $W$ is the width of the contacts and $d_\textrm{ij}$ and $w$ are the length and width of the channel, respectively. 
Due to the difficulty to fullfil these requirements, spin detection in the local configuration has been demonstrated in rare cases only.\cite{Ando2010, Ciorga2011, Nakane2010, Sasaki2011a, Althammer2012, Kasahara2012}

\begin{figure}[htbp!]
\includegraphics*[width=8cm]{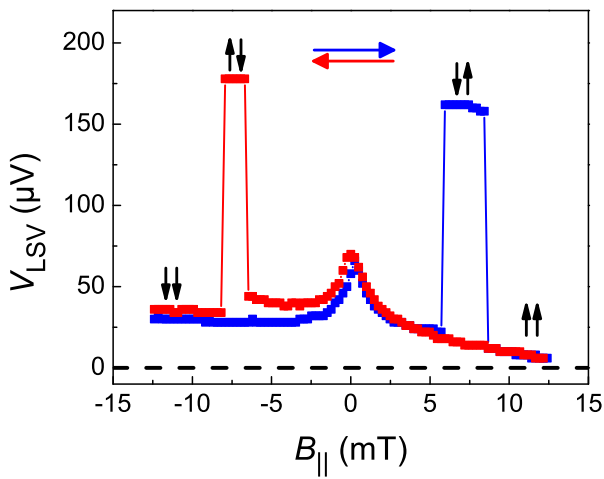}
\caption{Local spin signal in the spin valve geometry as a function of an in-plane magnetic field $B_{||}$ applied along the stripes. The peak around 0~T is induced by dynamic nuclear polarization.\cite{Salis2009b} The dashed line indicates the base line at 372.7~mV.}
\label{fig4}
\end{figure}

\begin{figure}[htbp!]
\includegraphics*[width=8cm]{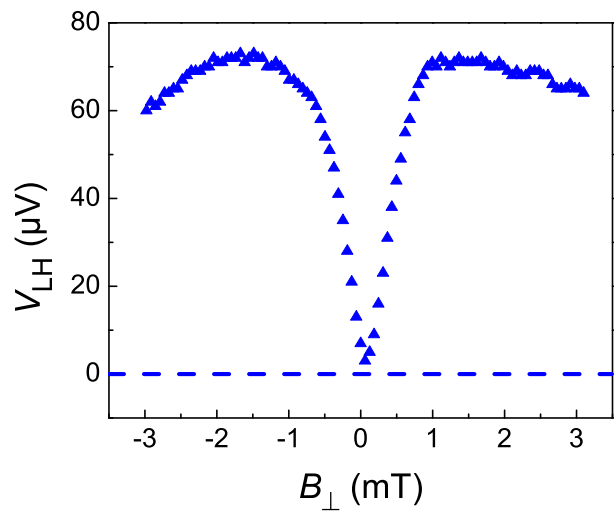}
\caption{Local voltage $V_\textrm{LH}$ as a function of the external out-of-plane magnetic field $B_\perp$ in the Hanle geometry measured at 40~K at a bias of $400~\mu$A. The dashed line indicates the base line at 548.38~mV.}
\label{fig5}
\end{figure}

Local spin valve (LSV) measurements have been performed in the same manner as described above for the non-local case except for the change in the contact configuration. Here, a charge current ($I_\textrm{L}$) flows between contacts~2 and 3 and the spin valve signal is measured as a voltage ($V_\textrm{L}$) between the same contacts [see Fig.~\ref{fig1}(b)]. As shown in Fig.~\ref{fig4}, the expected voltage jumps are clearly resolved and coincide 
 with those observed in the non-local configuration (cf.\ Fig.~\ref{fig2}). This observation provides clear evidence for local spin valve operation obtained in the Co$_2$FeSi/GaAs hybrid system. The observed Hanle curve for the local configuration, shown in Fig.~\ref{fig5}, supports our conclusion. Note that the slightly smaller linewidth as compared to the NL Hanle measurement at the same temperature (cf.\ Fig.~\ref{fig3}) indicates a somewhat larger spin lifetime.

The MR ratio for the samples under investigation is estimated to be $\Delta R/R_\textrm{P}\approx 0.03\%$ (cf.\ Fig.~\ref{fig4}), where the contact resistances were subtracted from the LSV curves by using data obtained by 3-terminal measurements.\cite{Ciorga2011} According to a theoretical estimate for the lateral geometry,\cite{Fert2001} which takes into account a spin dependent interface resistance at the ferromagnet/semiconductor interface, a MR ratio of 0.05\% is expected for our device, in reasonable agreement with our measured value. The relatively small MR ratio reflects the fact that the actual device parameters do not satisfy the condition expressed by Eq.~(\ref{window}).
The ratio $r^*_\textrm{b}/r_\textrm{N}\approx 32$ for the investigated devices is outside the optimal window, which for these samples is $1.6\ll r^*_b/r_N\ll 6.6$. Lowering the temperature below 40~K did not improve the MR ratio significantly.

The spin signal in the local configuration ($\Delta V_{\textrm{LSV}}=140~\mu$V) is larger than that in the non-local one ($\Delta V_{\textrm{NLSV}}=8~\mu$V) by a factor of about 18. This observation is in contrast to a one-dimensional spin diffusion model, where an expected ratio of $\Delta V_{\textrm{LSV}}/\Delta V_{\textrm{NLSV}}=2$ has been experimentally verified.\cite{Jedema2003} A deviation from this factor $2$ has been observed previously in FM/semiconductor hybrid systems and explained by an increase of the spin diffusion length $\lambda_{\textrm{sf}}$ in the local case due to the electric field in the semiconductor,\cite{Sasaki2011a} possibly due to an increase of $\tau_\textrm{s}$ as shown above.
Similar to the NL case (cf.\ inset of Fig.~\ref{fig2}), we measured the dependence of $\Delta V_{\textrm{LSV}}$ on the separation $d_{\textrm{ij}}$ and obtained indeed an increased spin diffusion length of $\lambda_{\textrm{sf}}\approx11~\mu$m.
Taking into account the different spin diffusion lengths, the ratio $\Delta V_{\textrm{LSV}}/\Delta V_{\textrm{NLSV}}$ has to be corrected by a factor of $4$ due to the different values of $\lambda_{\textrm{sf}}\times \exp(-d_\textrm{ij}/\lambda_{\textrm{sf}})$ in Eq.~(\ref{sv}). As a result, the corrected ratio $\Delta V_{\textrm{LSV}}/\Delta V_{\textrm{NLSV}}=18/4=4.5$ is still about a factor of $2$ larger than expected according to the theory for a one-dimensional spin diffusion model.\cite{Fert1996, Jedema2003} 
This remaining enhancement of the local spin valve signal might be related to the predicted half-metallic nature of Co$_2$FeSi in its \textit{L}2$_1$ phase. More precisely, the large spin polarization at the Fermi energy in Co$_2$FeSi may lead to a comparatively large spin detection efficiency in the local spin valve configuration.

We have presented unambiguous evidence for all-electrical spin injection and detection in the local and non-local configuration in the Co$_2$FeSi/GaAs hybrid system. The obtained magnetoresistance for the local spin valve configuration is found to be in accordance with the calculated estimate for (non-optimized) device parameters. The enhanced LSV signal with respect to the NL one suggests that the large spin polarization of Co$_2$FeSi in the well-ordered \textit{L}2$_1$ phase is advantageous for actual spintronic devices.


%
%

%

\begin{acknowledgments}
We gratefully acknowledge the technical support by Walid Anders and Angela Riedel and the critical reading of the manuscript by Alberto Hern\'andez-M\'inguez.
\end{acknowledgments}

\bibliographystyle{aipnum4-1}
\bibliography{Localnonlocal}

\begin{thebibliography}{31}%
\makeatletter
\providecommand \@ifxundefined [1]{%
 \@ifx{#1\undefined}
}%
\providecommand \@ifnum [1]{%
 \ifnum #1\expandafter \@firstoftwo
 \else \expandafter \@secondoftwo
 \fi
}%
\providecommand \@ifx [1]{%
 \ifx #1\expandafter \@firstoftwo
 \else \expandafter \@secondoftwo
 \fi
}%
\providecommand \natexlab [1]{#1}%
\providecommand \enquote  [1]{``#1''}%
\providecommand \bibnamefont  [1]{#1}%
\providecommand \bibfnamefont [1]{#1}%
\providecommand \citenamefont [1]{#1}%
\providecommand \href@noop [0]{\@secondoftwo}%
\providecommand \href [0]{\begingroup \@sanitize@url \@href}%
\providecommand \@href[1]{\@@startlink{#1}\@@href}%
\providecommand \@@href[1]{\endgroup#1\@@endlink}%
\providecommand \@sanitize@url [0]{\catcode `\\12\catcode `\$12\catcode
  `\&12\catcode `\#12\catcode `\^12\catcode `\_12\catcode `\%12\relax}%
\providecommand \@@startlink[1]{}%
\providecommand \@@endlink[0]{}%
\providecommand \url  [0]{\begingroup\@sanitize@url \@url }%
\providecommand \@url [1]{\endgroup\@href {#1}{\urlprefix }}%
\providecommand \urlprefix  [0]{URL }%
\providecommand \Eprint [0]{\href }%
\providecommand \doibase [0]{http://dx.doi.org/}%
\providecommand \selectlanguage [0]{\@gobble}%
\providecommand \bibinfo  [0]{\@secondoftwo}%
\providecommand \bibfield  [0]{\@secondoftwo}%
\providecommand \translation [1]{[#1]}%
\providecommand \BibitemOpen [0]{}%
\providecommand \bibitemStop [0]{}%
\providecommand \bibitemNoStop [0]{.\EOS\space}%
\providecommand \EOS [0]{\spacefactor3000\relax}%
\providecommand \BibitemShut  [1]{\csname bibitem#1\endcsname}%
\let\auto@bib@innerbib\@empty
\bibitem [{\citenamefont {Datta}\ and\ \citenamefont
  {Das}(1989)}]{DattaDas1989}%
  \BibitemOpen
  \bibfield  {author} {\bibinfo {author} {\bibfnamefont {S.}~\bibnamefont
  {Datta}}\ and\ \bibinfo {author} {\bibfnamefont {B.}~\bibnamefont {Das}},\
  }\href@noop {} {\bibfield  {journal} {\bibinfo  {journal} {Appl. Phys.
  Lett.}\ }\textbf {\bibinfo {volume} {56}},\ \bibinfo {pages} {665} (\bibinfo
  {year} {1989})}\BibitemShut {NoStop}%
\bibitem [{\citenamefont {Sugahara}\ and\ \citenamefont
  {Tanaka}(2004)}]{Sugahara2004}%
  \BibitemOpen
  \bibfield  {author} {\bibinfo {author} {\bibfnamefont {S.}~\bibnamefont
  {Sugahara}}\ and\ \bibinfo {author} {\bibfnamefont {M.}~\bibnamefont
  {Tanaka}},\ }\href {\doibase 10.1063/1.1689403} {\bibfield  {journal}
  {\bibinfo  {journal} {Appl. Phys. Lett.}\ }\textbf {\bibinfo {volume} {84}},\
  \bibinfo {pages} {2307} (\bibinfo {year} {2004})}\BibitemShut {NoStop}%
\bibitem [{\citenamefont {Johnson}\ and\ \citenamefont
  {Silsbee}(1985)}]{Johnson1985}%
  \BibitemOpen
  \bibfield  {author} {\bibinfo {author} {\bibfnamefont {M.}~\bibnamefont
  {Johnson}}\ and\ \bibinfo {author} {\bibfnamefont {R.~H.}\ \bibnamefont
  {Silsbee}},\ }\href {\doibase 10.1103/PhysRevLett.55.1790} {\bibfield
  {journal} {\bibinfo  {journal} {Phys. Rev. Lett.}\ }\textbf {\bibinfo
  {volume} {55}},\ \bibinfo {pages} {1790} (\bibinfo {year}
  {1985})}\BibitemShut {NoStop}%
\bibitem [{\citenamefont {Jedema}, \citenamefont {Filip},\ and\ \citenamefont
  {van Wees}(2001)}]{Jedema2001}%
  \BibitemOpen
  \bibfield  {author} {\bibinfo {author} {\bibfnamefont {F.~J.}\ \bibnamefont
  {Jedema}}, \bibinfo {author} {\bibfnamefont {A.~T.}\ \bibnamefont {Filip}}, \
  and\ \bibinfo {author} {\bibfnamefont {B.~J.}\ \bibnamefont {van Wees}},\
  }\href {http://dx.doi.org/10.1038/35066533} {\bibfield  {journal} {\bibinfo
  {journal} {Nature}\ }\textbf {\bibinfo {volume} {410}},\ \bibinfo {pages}
  {345} (\bibinfo {year} {2001})}\BibitemShut {NoStop}%
\bibitem [{\citenamefont {Tang}\ \emph {et~al.}(2002)\citenamefont {Tang} \emph
  {et~al.}}]{Tang2002}%
  \BibitemOpen
  \bibfield  {author} {\bibinfo {author} {\bibfnamefont {H.~X.}\ \bibnamefont
  {Tang}} \emph {et~al.},\ }\href@noop {} {\emph {\bibinfo {title}
  {Semicondoctor Spintronics and Quantum Computation}}},\ edited by\ \bibinfo
  {editor} {\bibfnamefont {D.~D.}\ \bibnamefont {Awschalom}}, \bibinfo {editor}
  {\bibfnamefont {D.}~\bibnamefont {Loss}}, \ and\ \bibinfo {editor}
  {\bibfnamefont {N.}~\bibnamefont {Samarth}}\ (\bibinfo  {publisher}
  {Springer},\ \bibinfo {year} {2002})\BibitemShut {NoStop}%
\bibitem [{\citenamefont {Wurmehl}\ \emph {et~al.}(2005)\citenamefont
  {Wurmehl}, \citenamefont {Fecher}, \citenamefont {Kandpal}, \citenamefont
  {Ksenofontov}, \citenamefont {Felser}, \citenamefont {Lin},\ and\
  \citenamefont {Morais}}]{Wurmehl2005}%
  \BibitemOpen
  \bibfield  {author} {\bibinfo {author} {\bibfnamefont {S.}~\bibnamefont
  {Wurmehl}}, \bibinfo {author} {\bibfnamefont {G.~H.}\ \bibnamefont {Fecher}},
  \bibinfo {author} {\bibfnamefont {H.~C.}\ \bibnamefont {Kandpal}}, \bibinfo
  {author} {\bibfnamefont {V.}~\bibnamefont {Ksenofontov}}, \bibinfo {author}
  {\bibfnamefont {C.}~\bibnamefont {Felser}}, \bibinfo {author} {\bibfnamefont
  {H.-J.}\ \bibnamefont {Lin}}, \ and\ \bibinfo {author} {\bibfnamefont
  {J.}~\bibnamefont {Morais}},\ }\href {\doibase 10.1103/PhysRevB.72.184434}
  {\bibfield  {journal} {\bibinfo  {journal} {Phys. Rev. B}\ }\textbf {\bibinfo
  {volume} {72}},\ \bibinfo {pages} {184434} (\bibinfo {year}
  {2005})}\BibitemShut {NoStop}%
\bibitem [{\citenamefont {Bruski}\ \emph {et~al.}(2011)\citenamefont {Bruski},
  \citenamefont {Erwin}, \citenamefont {Ramsteiner}, \citenamefont {Brandt},
  \citenamefont {Friedland}, \citenamefont {Farshchi}, \citenamefont
  {Herfort},\ and\ \citenamefont {Riechert}}]{Bruski2011}%
  \BibitemOpen
  \bibfield  {author} {\bibinfo {author} {\bibfnamefont {P.}~\bibnamefont
  {Bruski}}, \bibinfo {author} {\bibfnamefont {S.~C.}\ \bibnamefont {Erwin}},
  \bibinfo {author} {\bibfnamefont {M.}~\bibnamefont {Ramsteiner}}, \bibinfo
  {author} {\bibfnamefont {O.}~\bibnamefont {Brandt}}, \bibinfo {author}
  {\bibfnamefont {K.-J.}\ \bibnamefont {Friedland}}, \bibinfo {author}
  {\bibfnamefont {R.}~\bibnamefont {Farshchi}}, \bibinfo {author}
  {\bibfnamefont {J.}~\bibnamefont {Herfort}}, \ and\ \bibinfo {author}
  {\bibfnamefont {H.}~\bibnamefont {Riechert}},\ }\href {\doibase
  10.1103/PhysRevB.83.140409} {\bibfield  {journal} {\bibinfo  {journal} {Phys.
  Rev. B}\ }\textbf {\bibinfo {volume} {83}},\ \bibinfo {pages} {140409(R)}
  (\bibinfo {year} {2011})}\BibitemShut {NoStop}%
\bibitem [{\citenamefont {Hashimoto}\ \emph
  {et~al.}(2005{\natexlab{a}})\citenamefont {Hashimoto}, \citenamefont
  {Herfort}, \citenamefont {Sch\"{o}nherr},\ and\ \citenamefont
  {Ploog}}]{Hashimoto2005}%
  \BibitemOpen
  \bibfield  {author} {\bibinfo {author} {\bibfnamefont {M.}~\bibnamefont
  {Hashimoto}}, \bibinfo {author} {\bibfnamefont {J.}~\bibnamefont {Herfort}},
  \bibinfo {author} {\bibfnamefont {H.-P.}\ \bibnamefont {Sch\"{o}nherr}}, \
  and\ \bibinfo {author} {\bibfnamefont {K.~H.}\ \bibnamefont {Ploog}},\ }\href
  {\doibase 10.1063/1.2041836} {\bibfield  {journal} {\bibinfo  {journal}
  {Appl. Phys. Lett.}\ }\textbf {\bibinfo {volume} {87}},\ \bibinfo {eid}
  {102506} (\bibinfo {year} {2005}{\natexlab{a}})}\BibitemShut {NoStop}%
\bibitem [{\citenamefont {Ramsteiner}\ \emph {et~al.}(2008)\citenamefont
  {Ramsteiner}, \citenamefont {Brandt}, \citenamefont {Flissikowski},
  \citenamefont {Grahn}, \citenamefont {Hashimoto}, \citenamefont {Herfort},\
  and\ \citenamefont {Kostial}}]{Ramsteiner2008}%
  \BibitemOpen
  \bibfield  {author} {\bibinfo {author} {\bibfnamefont {M.}~\bibnamefont
  {Ramsteiner}}, \bibinfo {author} {\bibfnamefont {O.}~\bibnamefont {Brandt}},
  \bibinfo {author} {\bibfnamefont {T.}~\bibnamefont {Flissikowski}}, \bibinfo
  {author} {\bibfnamefont {H.~T.}\ \bibnamefont {Grahn}}, \bibinfo {author}
  {\bibfnamefont {M.}~\bibnamefont {Hashimoto}}, \bibinfo {author}
  {\bibfnamefont {J.}~\bibnamefont {Herfort}}, \ and\ \bibinfo {author}
  {\bibfnamefont {H.}~\bibnamefont {Kostial}},\ }\href {\doibase
  10.1103/PhysRevB.78.121303} {\bibfield  {journal} {\bibinfo  {journal} {Phys.
  Rev. B}\ }\textbf {\bibinfo {volume} {78}},\ \bibinfo {pages} {121303}
  (\bibinfo {year} {2008})}\BibitemShut {NoStop}%
\bibitem [{\citenamefont {Hashimoto}\ \emph
  {et~al.}(2005{\natexlab{b}})\citenamefont {Hashimoto}, \citenamefont
  {Herfort}, \citenamefont {Sch\"{o}nherr},\ and\ \citenamefont
  {Ploog}}]{Hashimoto2005a}%
  \BibitemOpen
  \bibfield  {author} {\bibinfo {author} {\bibfnamefont {M.}~\bibnamefont
  {Hashimoto}}, \bibinfo {author} {\bibfnamefont {J.}~\bibnamefont {Herfort}},
  \bibinfo {author} {\bibfnamefont {H.-P.}\ \bibnamefont {Sch\"{o}nherr}}, \
  and\ \bibinfo {author} {\bibfnamefont {K.~H.}\ \bibnamefont {Ploog}},\ }\href
  {\doibase 10.1063/1.2136213} {\bibfield  {journal} {\bibinfo  {journal} {J.
  Appl. Phys.}\ }\textbf {\bibinfo {volume} {98}},\ \bibinfo {eid} {104902}
  (\bibinfo {year} {2005}{\natexlab{b}})}\BibitemShut {NoStop}%
\bibitem [{\citenamefont {Hashimoto}\ \emph {et~al.}(2007)\citenamefont
  {Hashimoto}, \citenamefont {Trampert}, \citenamefont {Herfort},\ and\
  \citenamefont {Ploog}}]{Hashimoto2007}%
  \BibitemOpen
  \bibfield  {author} {\bibinfo {author} {\bibfnamefont {M.}~\bibnamefont
  {Hashimoto}}, \bibinfo {author} {\bibfnamefont {A.}~\bibnamefont {Trampert}},
  \bibinfo {author} {\bibfnamefont {J.}~\bibnamefont {Herfort}}, \ and\
  \bibinfo {author} {\bibfnamefont {K.~H.}\ \bibnamefont {Ploog}},\ }\href@noop
  {} {\bibfield  {journal} {\bibinfo  {journal} {J. Vac. Sci. Technol. B}\
  }\textbf {\bibinfo {volume} {25}},\ \bibinfo {pages} {1453} (\bibinfo {year}
  {2007})}\BibitemShut {NoStop}%
\bibitem [{\citenamefont {Kasahara}\ \emph {et~al.}(2012)\citenamefont
  {Kasahara}, \citenamefont {Baba}, \citenamefont {Yamane}, \citenamefont
  {Ando}, \citenamefont {Yamada}, \citenamefont {Hoshi}, \citenamefont
  {Sawano}, \citenamefont {Miyao},\ and\ \citenamefont
  {Hamaya}}]{Kasahara2012}%
  \BibitemOpen
  \bibfield  {author} {\bibinfo {author} {\bibfnamefont {K.}~\bibnamefont
  {Kasahara}}, \bibinfo {author} {\bibfnamefont {Y.}~\bibnamefont {Baba}},
  \bibinfo {author} {\bibfnamefont {K.}~\bibnamefont {Yamane}}, \bibinfo
  {author} {\bibfnamefont {Y.}~\bibnamefont {Ando}}, \bibinfo {author}
  {\bibfnamefont {S.}~\bibnamefont {Yamada}}, \bibinfo {author} {\bibfnamefont
  {Y.}~\bibnamefont {Hoshi}}, \bibinfo {author} {\bibfnamefont
  {K.}~\bibnamefont {Sawano}}, \bibinfo {author} {\bibfnamefont
  {M.}~\bibnamefont {Miyao}}, \ and\ \bibinfo {author} {\bibfnamefont
  {K.}~\bibnamefont {Hamaya}},\ }\href {\doibase 10.1063/1.3670985} {\bibfield
  {journal} {\bibinfo  {journal} {J. Appl. Phys.}\ }\textbf {\bibinfo {volume}
  {111}},\ \bibinfo {eid} {07C503} (\bibinfo {year} {2012})}\BibitemShut
  {NoStop}%
\bibitem [{\citenamefont {J\"onsson-\r{A}kerman}\ \emph
  {et~al.}(2000)\citenamefont {J\"onsson-\r{A}kerman}, \citenamefont
  {Escudero}, \citenamefont {Leighton}, \citenamefont {Kim}, \citenamefont
  {Schuller},\ and\ \citenamefont {Rabson}}]{Joensson2000}%
  \BibitemOpen
  \bibfield  {author} {\bibinfo {author} {\bibfnamefont {B.~J.}\ \bibnamefont
  {J\"onsson-\r{A}kerman}}, \bibinfo {author} {\bibfnamefont {R.}~\bibnamefont
  {Escudero}}, \bibinfo {author} {\bibfnamefont {C.}~\bibnamefont {Leighton}},
  \bibinfo {author} {\bibfnamefont {S.}~\bibnamefont {Kim}}, \bibinfo {author}
  {\bibfnamefont {I.~K.}\ \bibnamefont {Schuller}}, \ and\ \bibinfo {author}
  {\bibfnamefont {D.~A.}\ \bibnamefont {Rabson}},\ }\href {\doibase
  DOI:10.1063/1.1310633} {\bibfield  {journal} {\bibinfo  {journal} {Appl.
  Phys. Lett.}\ }\textbf {\bibinfo {volume} {77}},\ \bibinfo {pages} {1870}
  (\bibinfo {year} {2000})}\BibitemShut {NoStop}%
\bibitem [{\citenamefont {Salis}, \citenamefont {Fuhrer},\ and\ \citenamefont
  {Alvarado}(2009)}]{Salis2009b}%
  \BibitemOpen
  \bibfield  {author} {\bibinfo {author} {\bibfnamefont {G.}~\bibnamefont
  {Salis}}, \bibinfo {author} {\bibfnamefont {A.}~\bibnamefont {Fuhrer}}, \
  and\ \bibinfo {author} {\bibfnamefont {S.~F.}\ \bibnamefont {Alvarado}},\
  }\href {\doibase 10.1103/PhysRevB.80.115332} {\bibfield  {journal} {\bibinfo
  {journal} {Phys. Rev. B}\ }\textbf {\bibinfo {volume} {80}},\ \bibinfo
  {pages} {115332} (\bibinfo {year} {2009})}\BibitemShut {NoStop}%
\bibitem [{\citenamefont {Johnson}(1993)}]{Johnson1993}%
  \BibitemOpen
  \bibfield  {author} {\bibinfo {author} {\bibfnamefont {M.}~\bibnamefont
  {Johnson}},\ }\href {\doibase 10.1103/PhysRevLett.70.2142} {\bibfield
  {journal} {\bibinfo  {journal} {Phys. Rev. Lett.}\ }\textbf {\bibinfo
  {volume} {70}},\ \bibinfo {pages} {2142} (\bibinfo {year}
  {1993})}\BibitemShut {NoStop}%
\bibitem [{\citenamefont {Jedema}\ \emph {et~al.}(2003)\citenamefont {Jedema},
  \citenamefont {Nijboer}, \citenamefont {Filip},\ and\ \citenamefont {van
  Wees}}]{Jedema2003}%
  \BibitemOpen
  \bibfield  {author} {\bibinfo {author} {\bibfnamefont {F.~J.}\ \bibnamefont
  {Jedema}}, \bibinfo {author} {\bibfnamefont {M.~S.}\ \bibnamefont {Nijboer}},
  \bibinfo {author} {\bibfnamefont {A.~T.}\ \bibnamefont {Filip}}, \ and\
  \bibinfo {author} {\bibfnamefont {B.~J.}\ \bibnamefont {van Wees}},\ }\href
  {\doibase 10.1103/PhysRevB.67.085319} {\bibfield  {journal} {\bibinfo
  {journal} {Phys. Rev. B}\ }\textbf {\bibinfo {volume} {67}},\ \bibinfo
  {pages} {085319} (\bibinfo {year} {2003})}\BibitemShut {NoStop}%
\bibitem [{\citenamefont {Fabian}\ \emph {et~al.}(2007)\citenamefont {Fabian},
  \citenamefont {Matos-Abiaguea}, \citenamefont {Ertlera}, \citenamefont
  {Stano},\ and\ \citenamefont {\v{Z}uti\'{c}}}]{Fabian2007}%
  \BibitemOpen
  \bibfield  {author} {\bibinfo {author} {\bibfnamefont {J.}~\bibnamefont
  {Fabian}}, \bibinfo {author} {\bibfnamefont {A.}~\bibnamefont
  {Matos-Abiaguea}}, \bibinfo {author} {\bibfnamefont {C.}~\bibnamefont
  {Ertlera}}, \bibinfo {author} {\bibfnamefont {P.}~\bibnamefont {Stano}}, \
  and\ \bibinfo {author} {\bibfnamefont {I.}~\bibnamefont {\v{Z}uti\'{c}}},\
  }\href@noop {} {\bibfield  {journal} {\bibinfo  {journal} {Acta Physica
  Slovaca}\ }\textbf {\bibinfo {volume} {57}},\ \bibinfo {pages} {565}
  (\bibinfo {year} {2007})}\BibitemShut {NoStop}%
\bibitem [{\citenamefont {Lou}\ \emph {et~al.}(2007)\citenamefont {Lou},
  \citenamefont {Adelmann}, \citenamefont {Crooker}, \citenamefont {Garlid},
  \citenamefont {Zhang}, \citenamefont {Reddy}, \citenamefont {Flexner},
  \citenamefont {Palmstr\o{}m},\ and\ \citenamefont {Crowell}}]{Lou2007}%
  \BibitemOpen
  \bibfield  {author} {\bibinfo {author} {\bibfnamefont {X.}~\bibnamefont
  {Lou}}, \bibinfo {author} {\bibfnamefont {C.}~\bibnamefont {Adelmann}},
  \bibinfo {author} {\bibfnamefont {S.~A.}\ \bibnamefont {Crooker}}, \bibinfo
  {author} {\bibfnamefont {E.~S.}\ \bibnamefont {Garlid}}, \bibinfo {author}
  {\bibfnamefont {J.}~\bibnamefont {Zhang}}, \bibinfo {author} {\bibfnamefont
  {K.~S.~M.}\ \bibnamefont {Reddy}}, \bibinfo {author} {\bibfnamefont {S.~D.}\
  \bibnamefont {Flexner}}, \bibinfo {author} {\bibfnamefont {C.~J.}\
  \bibnamefont {Palmstr\o{}m}}, \ and\ \bibinfo {author} {\bibfnamefont
  {P.~A.}\ \bibnamefont {Crowell}},\ }\href
  {http://dx.doi.org/10.1038/nphys543} {\bibfield  {journal} {\bibinfo
  {journal} {Nat. Phys.}\ }\textbf {\bibinfo {volume} {3}},\ \bibinfo {pages}
  {197} (\bibinfo {year} {2007})}\BibitemShut {NoStop}%
\bibitem [{\citenamefont {Ciorga}\ \emph {et~al.}(2009)\citenamefont {Ciorga},
  \citenamefont {Einwanger}, \citenamefont {Wurstbauer}, \citenamefont {Schuh},
  \citenamefont {Wegscheider},\ and\ \citenamefont {Weiss}}]{Ciorga2009}%
  \BibitemOpen
  \bibfield  {author} {\bibinfo {author} {\bibfnamefont {M.}~\bibnamefont
  {Ciorga}}, \bibinfo {author} {\bibfnamefont {A.}~\bibnamefont {Einwanger}},
  \bibinfo {author} {\bibfnamefont {U.}~\bibnamefont {Wurstbauer}}, \bibinfo
  {author} {\bibfnamefont {D.}~\bibnamefont {Schuh}}, \bibinfo {author}
  {\bibfnamefont {W.}~\bibnamefont {Wegscheider}}, \ and\ \bibinfo {author}
  {\bibfnamefont {D.}~\bibnamefont {Weiss}},\ }\href {\doibase
  10.1103/PhysRevB.79.165321} {\bibfield  {journal} {\bibinfo  {journal} {Phys.
  Rev. B}\ }\textbf {\bibinfo {volume} {79}},\ \bibinfo {pages} {165321}
  (\bibinfo {year} {2009})}\BibitemShut {NoStop}%
\bibitem [{\citenamefont {Salis}\ \emph {et~al.}(2010)\citenamefont {Salis},
  \citenamefont {Fuhrer}, \citenamefont {Schlittler}, \citenamefont {Gross},\
  and\ \citenamefont {Alvarado}}]{Salis2010}%
  \BibitemOpen
  \bibfield  {author} {\bibinfo {author} {\bibfnamefont {G.}~\bibnamefont
  {Salis}}, \bibinfo {author} {\bibfnamefont {A.}~\bibnamefont {Fuhrer}},
  \bibinfo {author} {\bibfnamefont {R.~R.}\ \bibnamefont {Schlittler}},
  \bibinfo {author} {\bibfnamefont {L.}~\bibnamefont {Gross}}, \ and\ \bibinfo
  {author} {\bibfnamefont {S.~F.}\ \bibnamefont {Alvarado}},\ }\href {\doibase
  10.1103/PhysRevB.81.205323} {\bibfield  {journal} {\bibinfo  {journal} {Phys.
  Rev. B}\ }\textbf {\bibinfo {volume} {81}},\ \bibinfo {pages} {205323}
  (\bibinfo {year} {2010})}\BibitemShut {NoStop}%
\bibitem [{\citenamefont {Jedema}\ \emph {et~al.}(2002)\citenamefont {Jedema},
  \citenamefont {Heersche}, \citenamefont {Filip}, \citenamefont {Baselmans},\
  and\ \citenamefont {van Wees}}]{Jedema2002}%
  \BibitemOpen
  \bibfield  {author} {\bibinfo {author} {\bibfnamefont {F.~J.}\ \bibnamefont
  {Jedema}}, \bibinfo {author} {\bibfnamefont {H.~B.}\ \bibnamefont
  {Heersche}}, \bibinfo {author} {\bibfnamefont {A.~T.}\ \bibnamefont {Filip}},
  \bibinfo {author} {\bibfnamefont {J.~J.~A.}\ \bibnamefont {Baselmans}}, \
  and\ \bibinfo {author} {\bibfnamefont {B.~J.}\ \bibnamefont {van Wees}},\
  }\href {http://dx.doi.org/10.1038/416713a} {\bibfield  {journal} {\bibinfo
  {journal} {Nature}\ }\textbf {\bibinfo {volume} {416}},\ \bibinfo {pages}
  {713} (\bibinfo {year} {2002})}\BibitemShut {NoStop}%
\bibitem [{\citenamefont {Fert}\ and\ \citenamefont
  {Jaffr\`es}(2001)}]{Fert2001}%
  \BibitemOpen
  \bibfield  {author} {\bibinfo {author} {\bibfnamefont {A.}~\bibnamefont
  {Fert}}\ and\ \bibinfo {author} {\bibfnamefont {H.}~\bibnamefont
  {Jaffr\`es}},\ }\href {\doibase 10.1103/PhysRevB.64.184420} {\bibfield
  {journal} {\bibinfo  {journal} {Phys. Rev. B}\ }\textbf {\bibinfo {volume}
  {64}},\ \bibinfo {pages} {184420} (\bibinfo {year} {2001})}\BibitemShut
  {NoStop}%
\bibitem [{\citenamefont {Dery}, \citenamefont {Cywi\ifmmode~\acute{n}\else
  \'{n}\fi{}ski},\ and\ \citenamefont {Sham}(2006)}]{Dery2006}%
  \BibitemOpen
  \bibfield  {author} {\bibinfo {author} {\bibfnamefont {H.}~\bibnamefont
  {Dery}}, \bibinfo {author} {\bibfnamefont {L.}~\bibnamefont
  {Cywi\ifmmode~\acute{n}\else \'{n}\fi{}ski}}, \ and\ \bibinfo {author}
  {\bibfnamefont {L.~J.}\ \bibnamefont {Sham}},\ }\href {\doibase
  10.1103/PhysRevB.73.041306} {\bibfield  {journal} {\bibinfo  {journal} {Phys.
  Rev. B}\ }\textbf {\bibinfo {volume} {73}},\ \bibinfo {pages} {041306}
  (\bibinfo {year} {2006})}\BibitemShut {NoStop}%
\bibitem [{\citenamefont {Fert}\ \emph {et~al.}(2007)\citenamefont {Fert},
  \citenamefont {George}, \citenamefont {Jaffres},\ and\ \citenamefont
  {Mattana}}]{Fert2007}%
  \BibitemOpen
  \bibfield  {author} {\bibinfo {author} {\bibfnamefont {A.}~\bibnamefont
  {Fert}}, \bibinfo {author} {\bibfnamefont {J.~M.}\ \bibnamefont {George}},
  \bibinfo {author} {\bibfnamefont {H.}~\bibnamefont {Jaffres}}, \ and\
  \bibinfo {author} {\bibfnamefont {R.}~\bibnamefont {Mattana}},\ }\href
  {\doibase 10.1109/TED.2007.894372} {\bibfield  {journal} {\bibinfo  {journal}
  {IEEE Transactions on Electron Devices}\ }\textbf {\bibinfo {volume} {54}},\
  \bibinfo {pages} {921} (\bibinfo {year} {2007})}\BibitemShut {NoStop}%
\bibitem [{\citenamefont {Schmidt}\ \emph {et~al.}(2000)\citenamefont
  {Schmidt}, \citenamefont {Ferrand}, \citenamefont {Molenkamp}, \citenamefont
  {Filip},\ and\ \citenamefont {van Wees}}]{Schmidt2000}%
  \BibitemOpen
  \bibfield  {author} {\bibinfo {author} {\bibfnamefont {G.}~\bibnamefont
  {Schmidt}}, \bibinfo {author} {\bibfnamefont {D.}~\bibnamefont {Ferrand}},
  \bibinfo {author} {\bibfnamefont {L.~W.}\ \bibnamefont {Molenkamp}}, \bibinfo
  {author} {\bibfnamefont {A.~T.}\ \bibnamefont {Filip}}, \ and\ \bibinfo
  {author} {\bibfnamefont {B.~J.}\ \bibnamefont {van Wees}},\ }\href {\doibase
  10.1103/PhysRevB.62.R4790} {\bibfield  {journal} {\bibinfo  {journal} {Phys.
  Rev. B}\ }\textbf {\bibinfo {volume} {62}},\ \bibinfo {pages} {R4790}
  (\bibinfo {year} {2000})}\BibitemShut {NoStop}%
\bibitem [{\citenamefont {Ando}\ \emph {et~al.}(2010)\citenamefont {Ando},
  \citenamefont {Kasahara}, \citenamefont {Yamane}, \citenamefont {Hamaya},
  \citenamefont {Sawano}, \citenamefont {Kimura},\ and\ \citenamefont
  {Miyao}}]{Ando2010}%
  \BibitemOpen
  \bibfield  {author} {\bibinfo {author} {\bibfnamefont {Y.}~\bibnamefont
  {Ando}}, \bibinfo {author} {\bibfnamefont {K.}~\bibnamefont {Kasahara}},
  \bibinfo {author} {\bibfnamefont {K.}~\bibnamefont {Yamane}}, \bibinfo
  {author} {\bibfnamefont {K.}~\bibnamefont {Hamaya}}, \bibinfo {author}
  {\bibfnamefont {K.}~\bibnamefont {Sawano}}, \bibinfo {author} {\bibfnamefont
  {T.}~\bibnamefont {Kimura}}, \ and\ \bibinfo {author} {\bibfnamefont
  {M.}~\bibnamefont {Miyao}},\ }\href {\doibase 10.1143/APEX.3.093001}
  {\bibfield  {journal} {\bibinfo  {journal} {Applied Physics Express}\
  }\textbf {\bibinfo {volume} {3}},\ \bibinfo {pages} {093001} (\bibinfo {year}
  {2010})}\BibitemShut {NoStop}%
\bibitem [{\citenamefont {Ciorga}\ \emph {et~al.}(2011)\citenamefont {Ciorga},
  \citenamefont {Wolf}, \citenamefont {Einwanger}, \citenamefont {Utz},
  \citenamefont {Schuh},\ and\ \citenamefont {Weiss}}]{Ciorga2011}%
  \BibitemOpen
  \bibfield  {author} {\bibinfo {author} {\bibfnamefont {M.}~\bibnamefont
  {Ciorga}}, \bibinfo {author} {\bibfnamefont {C.}~\bibnamefont {Wolf}},
  \bibinfo {author} {\bibfnamefont {A.}~\bibnamefont {Einwanger}}, \bibinfo
  {author} {\bibfnamefont {M.}~\bibnamefont {Utz}}, \bibinfo {author}
  {\bibfnamefont {D.}~\bibnamefont {Schuh}}, \ and\ \bibinfo {author}
  {\bibfnamefont {D.}~\bibnamefont {Weiss}},\ }\href {\doibase
  10.1063/1.3591397} {\bibfield  {journal} {\bibinfo  {journal} {AIP Advances}\
  }\textbf {\bibinfo {volume} {1}},\ \bibinfo {eid} {022113} (\bibinfo {year}
  {2011})}\BibitemShut {NoStop}%
\bibitem [{\citenamefont {Nakane}\ \emph {et~al.}(2010)\citenamefont {Nakane},
  \citenamefont {Harada}, \citenamefont {Sugiura},\ and\ \citenamefont
  {Tanaka}}]{Nakane2010}%
  \BibitemOpen
  \bibfield  {author} {\bibinfo {author} {\bibfnamefont {R.}~\bibnamefont
  {Nakane}}, \bibinfo {author} {\bibfnamefont {T.}~\bibnamefont {Harada}},
  \bibinfo {author} {\bibfnamefont {K.}~\bibnamefont {Sugiura}}, \ and\
  \bibinfo {author} {\bibfnamefont {M.}~\bibnamefont {Tanaka}},\ }\href
  {\doibase 10.1143/JJAP.49.113001} {\bibfield  {journal} {\bibinfo  {journal}
  {Jap. J. Appl. Phys.}\ }\textbf {\bibinfo {volume} {49}},\ \bibinfo {pages}
  {113001} (\bibinfo {year} {2010})}\BibitemShut {NoStop}%
\bibitem [{\citenamefont {Sasaki}\ \emph {et~al.}(2011)\citenamefont {Sasaki},
  \citenamefont {Oikawa}, \citenamefont {Suzuki}, \citenamefont {Shiraishi},
  \citenamefont {Suzuki},\ and\ \citenamefont {Noguchi}}]{Sasaki2011a}%
  \BibitemOpen
  \bibfield  {author} {\bibinfo {author} {\bibfnamefont {T.}~\bibnamefont
  {Sasaki}}, \bibinfo {author} {\bibfnamefont {T.}~\bibnamefont {Oikawa}},
  \bibinfo {author} {\bibfnamefont {T.}~\bibnamefont {Suzuki}}, \bibinfo
  {author} {\bibfnamefont {M.}~\bibnamefont {Shiraishi}}, \bibinfo {author}
  {\bibfnamefont {Y.}~\bibnamefont {Suzuki}}, \ and\ \bibinfo {author}
  {\bibfnamefont {K.}~\bibnamefont {Noguchi}},\ }\href {\doibase
  10.1063/1.3604010} {\bibfield  {journal} {\bibinfo  {journal} {Appl. Phys.
  Lett.}\ }\textbf {\bibinfo {volume} {98}},\ \bibinfo {eid} {262503} (\bibinfo
  {year} {2011})}\BibitemShut {NoStop}%
\bibitem [{\citenamefont {Althammer}\ \emph {et~al.}(2012)\citenamefont
  {Althammer}, \citenamefont {Karrer-M\"{u}ller}, \citenamefont {Goennenwein},
  \citenamefont {Opel},\ and\ \citenamefont {Gross}}]{Althammer2012}%
  \BibitemOpen
  \bibfield  {author} {\bibinfo {author} {\bibfnamefont {M.}~\bibnamefont
  {Althammer}}, \bibinfo {author} {\bibfnamefont {E.-M.}\ \bibnamefont
  {Karrer-M\"{u}ller}}, \bibinfo {author} {\bibfnamefont {S.~T.~B.}\
  \bibnamefont {Goennenwein}}, \bibinfo {author} {\bibfnamefont
  {M.}~\bibnamefont {Opel}}, \ and\ \bibinfo {author} {\bibfnamefont
  {R.}~\bibnamefont {Gross}},\ }\href {\doibase 10.1063/1.4747321} {\bibfield
  {journal} {\bibinfo  {journal} {Appl. Phys. Lett.}\ }\textbf {\bibinfo
  {volume} {101}},\ \bibinfo {eid} {082404} (\bibinfo {year}
  {2012})}\BibitemShut {NoStop}%
\bibitem [{\citenamefont {Fert}\ and\ \citenamefont {Lee}(1996)}]{Fert1996}%
  \BibitemOpen
  \bibfield  {author} {\bibinfo {author} {\bibfnamefont {A.}~\bibnamefont
  {Fert}}\ and\ \bibinfo {author} {\bibfnamefont {S.-F.}\ \bibnamefont {Lee}},\
  }\href {\doibase 10.1103/PhysRevB.53.6554} {\bibfield  {journal} {\bibinfo
  {journal} {Phys. Rev. B}\ }\textbf {\bibinfo {volume} {53}},\ \bibinfo
  {pages} {6554} (\bibinfo {year} {1996})}\BibitemShut {NoStop}%
\end{thebibliography}%

\end{document}